\documentclass[conference, 10pt]{IEEEtran}

\IEEEoverridecommandlockouts
\usepackage{cite}
\usepackage{amsmath,amssymb,amsfonts}
\usepackage{algorithmic,bm}
\usepackage{graphicx}
\usepackage{textcomp}
\usepackage{xcolor}
\usepackage{caption}
\usepackage{subfigure}
\usepackage{epstopdf}
\usepackage{multirow}
\usepackage{booktabs}
\usepackage{epsfig}
\usepackage{float}
\usepackage{url}
\usepackage{array}
\def\BibTeX{{\rm B\kern-.05em{\sc i\kern-.025em b}\kern-.08em
		T\kern-.1667em\lower.7ex\hbox{E}\kern-.125emX}}

\begin{document}
	
	%%%%%%%%%%%%%%%%%%%%%%%%%%%%%%%%%%%%%%%%%%%%%%%%%%%%%%%%%%%%%%%%%%%%%%%%%%%%%%%%%%%%%%%%%%%%%%%%%%%%%%%%%%%%
	\title{Knowledge-Driven Machine Learning: Concept, Model and Case Study on Channel Estimation}
	
	\author{
		\IEEEauthorblockN{
			Daofeng Li\IEEEauthorrefmark{1},
			Kaihe Deng\IEEEauthorrefmark{1},
			Ming Zhao\IEEEauthorrefmark{1},
			Sihai Zhang\IEEEauthorrefmark{1},
			Jinkang Zhu\IEEEauthorrefmark{2}}

		\IEEEauthorrefmark{1}Key Laboratory of Wireless-Optical Communications, \\
		University of Science \& Technology of China, Hefei, Anhui, P.R. China\\
		\IEEEauthorrefmark{2} PCNSS, University of Science \& Technology of China, Hefei, Anhui, P.R. China\\
		
		Email: \{df007,dengkh\}@mail.ustc.edu.cn, \{zhaoming, shzhang, jkzhu\}@ustc.edu.cn	
	}
	\maketitle
	\thispagestyle{empty}
	\begin{abstract}
		The power of big data and machine learning has been drastically demonstrated in many fields during the past twenty years which somehow leads to the vague even false understanding that the huge amount of precious human knowledge accumulated to date no longer seems to matter. 
		%But this is surely not true at least in machine learning.
		In this paper, we are pioneering to propose the knowledge-driven machine learning(KDML) model to exhibit that knowledge can play an important role in machine learning tasks.
		KDML takes advantage of domain knowledge to processes the input data by space transforming without any training which enable the space of input and the output data of the neural networks to be identical, so that we can simplify the machine learning network structure and reduce training costs significantly.
		Channel estimation problems considering the time selective and frequency selective fading in wireless communications are taken as a case study, where we choose least square(LS) and minimum mean-square error(MMSE) as knowledge module and Long Short Term Memory(LSTM) as learning module.
		The performance obtained by KDML channel estimator obviously outperforms that of knowledge processing or conventional machine learning, respectively.
		Our work sheds light on the new area of machine learning and knowledge processing.
	\end{abstract}
	
	\begin{IEEEkeywords}
		Knowledge-Driven, Machine Learning, Channel estimator, LSTM, Fine-tune
	\end{IEEEkeywords}
	
	\section{Introduction}
	\label{Intro}
	%Wireless communication is a field that has been extensively studied.
	%The new generation communication systems are supposed to provide high capacity, high data rates, and high mobility.
	
	%P1: wireless big data, AI, ML, with wireless comm.
	
	The prosperity of big data and machine learning may gush out of many researcher's expectations in the past decade.
	The data storage and computing capacity keep yet improving which can accommodate and utilize the explosively growing data traffic.
	Emerging technologies, such as cloud computing, the Internet of Things(IoT), artificial intelligence(AI) have been changing the whole life of human society.
	Machine learning (ML) has been already successfully applied into many fields, such as data mining\cite{Wlodarczak2015Multimedia}, natural language processing \cite{Deng2014Deep}, computer vision\cite{Krizhevsky2012ImageNet}, even wireless big data(WBD) in wireless communications\cite{QianSurvey}\cite{8375943}.

	However, ML still faces lots of challenges when combined with wireless communications. First of all, although machine learning models have strong nonlinear fitting ability, its performance will significantly decrease over different testing datasets due to over-fitting\cite{Lawrence2000Overfitting}, which makes it difficult to adapt to rapid changing wireless communication environments.  
	Besides, with the increasing complexity of ML, the requirements of data amount and computing ability also grow rapidly, which brings huge overhead of storage and computing.
	%But in reality, the storage and computing capacity of wireless communication equipments are quite limited.
	Furthermore, wireless communication systems often require high reliability, which is also agnostic because deep neural networks obtain high fitting power at the cost of low interpretability of their black-box representations at present\cite{Zhang2018Visual}.

	Confronted with the above challenges, many researchers have put forward new ML approaches to implement the combination of ML and wireless communications. 
	To enhance the generalization ability, transfer learning is proposed which focuses on the need for lifelong machine learning methods that retain and reuse previously learned knowledge\cite{Pan2010A}. 
	Likewise, meta-learning refers to the process of improving a learning algorithm over multiple learning episodes\cite{Hospedales2020Meta}. 
	As for the interpretability, the concept of wireless knowledge and knowledge entropy are addressed and wireless knowledge learning(WKL) is presented to improve the interpretability\cite{8594713}. 
	Nevertheless, neither transfer learning nor WKL has a clear interaction mechanism between wireless domain knowledge and neural networks.  
	%	They just replace the traditional algorithms with ML algorithms.
	Besides, these methods often need a huge training dataset and much higher complexity than traditional algorithms.
	Therefore, in this paper we propose the wireless knowledge-driven machine learning(KDML) model to combine wireless knowledge with ML. We treat traditional algorithms and models as knowledge and process the input data to ensure that the input space of learning module is the same as its output space.
	Thus, we can use a simpler learning structure to refine the performance of traditional algorithms with no obvious cost increase in terms of time complexity.

	When it comes to the case study of channel estimation, there are two categories of methods to apply ML algorithms. 
	One is constructing an end-to-end system to  simulate the wireless communication link by a specific machine learning network.
	The auto-encoder simulates one communication system over an AWGN channel, whose design is regarded as an end-to-end reconstruction that seeks to jointly optimize transmitter and receiver components together\cite{8054694}.
	Similarly, \cite{8214233} proposes a fine-tuned auto-encoder with a two-phase training strategy, which adopts transfer learning to overcome the channel gradient missing issue.
	The other takes advantages of the block structures of wireless communication system and constructs an independent channel estimator implemented by ML.
	%Besides, there are also many works studying channel estimation rather than simulating the full link communication. 	
	%\cite{8052521} considers signal detection and channel estimation in the Orthogonal Frequency Division Multiplexing (OFDM) system \cite{FlochCoded} and deep neural network is firstly trained by simulation data and then recover the transmitted data online.
	Multiple Layers Perceptron(MLP) is used to assist channel estimation and the channel estimator combines learning techniques with training symbols in preambles and pilots, and thus can track channel variations on-line\cite{8491068}.
	\cite{GRU} takes advantage of the sliding bidirectional recurrent neural network to estimate the time selective fading channel without any prior knowledge about the channel model.
	Yet, most ML based channel estimators completely ignore well-established traditional algorithms, such as LS and MMSE, which are well studied and summarized as domain knowledge in wireless communications. 
	Such domain knowledge can play an important role in ML, thus we design a two-steps fine-tuning channel estimator as the implementation of KDML to utilize domain knowledge.
	The estimation by traditional algorithms will be regarded as a time series and then put into the Long Short Term Memory(LSTM) network to extract the correlation hidden in the CSI.
%	Although there are rare works based on ML considering both time and frequency selective fading channel model, we consider both two fading characteristics to verify the effectiveness of KDML channel estimator
	%Besides, the channel environment is still more complicated than the channel model considered in previous works.
	%Time/frequency selective fading has already been well studied in wireless communication, but there are rare works based on ML considering these two fading characteristics.
	%Hence, we consider both two fading characteristics to support the high mobility and super-complicated environments.	
	
	The major contributions of this paper are as follows:
	\begin{itemize}	
		\item Knowledge-driven machine learning(KDML) is proposed which combines the knowledge processing with machine learning.
		KDML, to the best of our knowledge, is the pioneering work in the machine learning field.
		It uses domain knowledge to make sure the input space of ML is the same as its output space without any training which makes KDML more reliable than conventional ML. The network structure can be simplified and the training costs can be reduced. 
		\item Channel estimation problem considering the time selective and frequency selective fading is adopted as case study. 
		Least square(LS) and minimum mean-square error(MMSE) are chosen as knowledge and Long Short Term Memory(LSTM) as learning module.
		Extensive experiments verify that KDML channel estimator has much better performance than traditional algorithms or conventional ML while its time complexity is much lower than MMSE. 
		%and indicate that the performance of KDML outperforms that of separate LS, MMSE or separate LSTM.
	\end{itemize}

	\section{Knowledge-Driven Machine Learning}
	In this section, we introduce the data-driven ML at first. 
	Then, we propose the model of the knowledge-driven machine learning and discuss the differences between KDML, transfer learning and model-driven ML.
	
	\subsection{Data-Driven Machine Learning}
	%It is worth noting that KDML is based on the data-driven method but emphasizes the importance of knowledge.
	Pure or traditional data-driven ML model is shown in Fig.\ref{LM1}. 
	Let $\bm{x}$, $\bm{y}$, $\bm{\theta}$ denote the features, labels of dataset and parameters of the neural network, respectively. 
	The set of all possible values for the input, output and parameters is called the input space($\mathbb{X}$), the output space($\mathbb{Y}$) and the parameters space($\mathbf{\mathbf{\Theta}}$), respectively. 
	The tasks of data-driven models are to map instances of dataset from $\mathbb{X}$ to $\mathbb{Y}$ and find the global optimum solution in $\mathbf{\Theta}$ at the same time.
	It is not difficult to find that $\mathbb{X}$ is often completely disjoint with $\mathbb{Y}$ in most wireless communication cases. 
	Therefore, pure data-driven ML network is supposed to be optimized over a huge dataset without a mathematical model, which implies complex network structures and huge training costs.
	
	\subsection{Knowledge-Driven Machine Learning(KDML)}
	\label{proposed method}
	Considering the defects of data-driven ML, we propose the knowledge driven machine learning model which aims at taking advantage of wireless knowledge to simplify the ML network structure and reduce the training costs.
	It is worth noting that KDML is based on the data-driven method but emphasizes the importance of knowledge. Here, knowledge is the general term of the description, understanding, and cognition for all issues involved in a wide variety of domains\cite{8594713}.
	
	\begin{figure}[!t]
		\centering
		\subfigure[Machine Learning Model]{
			\begin{minipage}[b]{0.27\textwidth}
				\includegraphics[width=1\textwidth]{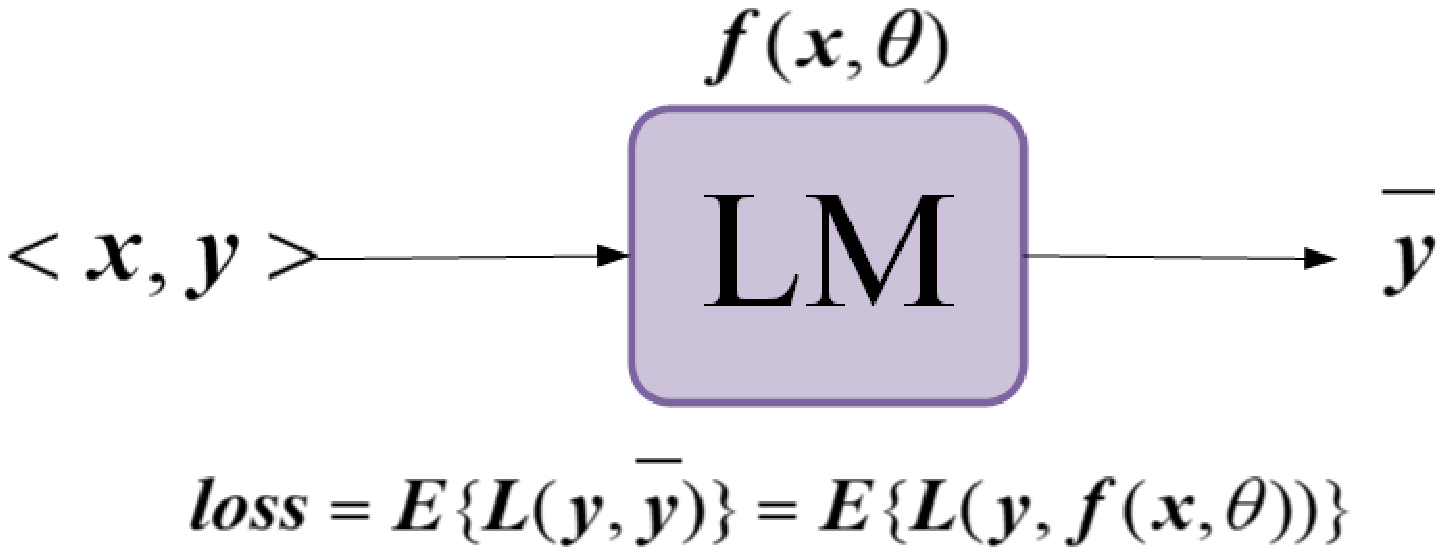}
			\end{minipage}
			\label{LM1}
		}
		\subfigure[Knowledge-Driven Machine Learning Model]{
			\begin{minipage}[b]{0.45\textwidth}
				\includegraphics[width=1\textwidth]{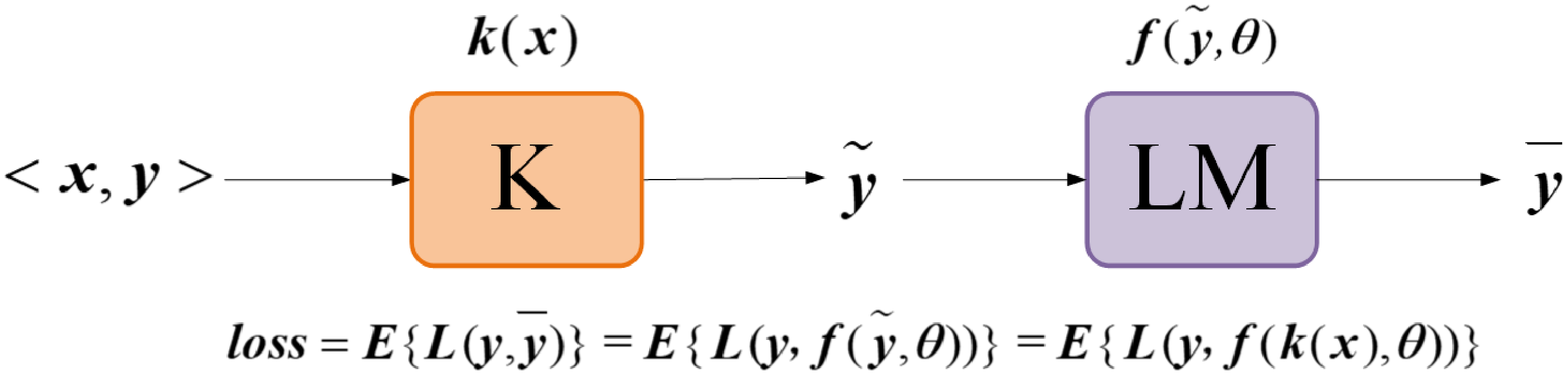}
			\end{minipage}
			\label{LM2}
		}
		\caption{Model Illustration of ML and KDML}
	\end{figure}

	The basic model of KDML is shown in Fig.\ref{LM2}. 
	%the knowledge-driven machine learning model(KDML) is proposed to combine the power of knowledge and data in this paper.
	Compared with data-driven models, a knowledge module denoted by $\bm{k}$ is added, whose functionality is to mine the commonalities over dataset based on knowledge. More specifically, it maps instances of dataset from $\mathbb{X}$ directly into $\mathbb{Y}$ as $\widetilde {\bm{y}}$ without any training.
	%After this module, the input space of ML is the same as the output space.
	And then, the learning module will further explore more specific and accurate information based on $\widetilde {\bm{y}}$. The objective function of KDML is given as :
	\begin{equation}
	\mathop{\arg\min}\limits_{\bm{\theta} \in \mathbf{\mathbf{\Theta}}} \left\|\bm{y}-f(\bm{k}(\bm{x}),\bm{\theta}) \right\|
	\end{equation}
	The most important characteristics of KDML is using  domain knowledge to rebuild learning tasks and make sure that the input space of ML is the same as the output space. 
	So, the number of input features of the neural network will decrease obviously and the training costs will be reduced correspondingly.
	Moreover, the interaction mechanism of the knowledge module and learning module is to treat the output of the former as the input of the latter, so that the learning module only considers searching for a global optimal solution without space transformation.

	\subsection{KDML versus Transfer Learning and MDML}

	There have already been some works on how to use domain knowledge to design an ML network for wireless communications, which can be broadly divided into two categories: transfer learning and model-driven ML(MDML). 
	Both methods have significant differences with KDML, as shown in TABLE \ref{KDML}.
	
	The goal of transfer learning is to choose a better initialization state for the ML network. 
	It trains the ML network over a huge dataset generated by domain knowledge at the first step. 
	Then the parameters obtained above will be used as initialization parameters for the new learning task. 
	Transfer learning uses knowledge to extend $\mathbb{X}$ and then reduce $\mathbf{\Theta}$, thus accelerates the speed of searching optimal solution. Therefore, transfer learning inevitably increases the cost of training. 
	Meanwhile, KDML lets the learning module no longer care about space transformation which may effectively reduce training costs.
	
		\begin{table}[t]
		\centering
		\caption{KDML versus other methods}
		\setlength{\tabcolsep}{1mm}
		\begin{tabular}{|c|c|c|}\hline
			\textbf{Methods}&\textbf{Objective Function}&\textbf{Additional operations}  \\ 		\hline
			Data-driven& \multirow{5}{*}{$   \mathop{\arg\min}\limits_{\bm{\theta} \in \mathbf{\mathbf{\Theta}}} \left\|\bm{y}-f(\bm{x},\bm{\theta}) \right\|$} & ---  \\ 		\cline{1-1} \cline{3-3} 
			Transfer Learning&  &Extend $\mathbb{X}$; Reduce $\mathbf{\mathbf{\Theta}}$\\  \cline{1-1} \cline{3-3} 
			\multirow{2}{*}	{Model-driven}& & NN Structure mimics\\ 
			& & traditional algorithms\\ \cline{1-3}
			\multirow{2}{*}	{KDML}&$   \mathop{\arg\min}\limits_{\bm{\theta} \in \mathbf{\mathbf{\Theta}}} \left\|\bm{y}-f(\bm{k}(\bm{x}),\bm{\theta}) \right\| $ &\multirow{2}{*}{ $k:\mathbb{X} \to \mathbb{Y}$ }\\ \hline
		\end{tabular}
		\label{KDML}
	\end{table}
	
	As for MDML, the main characteristics are the learning structure being constructed based on domain knowledge rather than huge volume of labeled data\cite{8715338}. 
	The learning structure will mimic the process of traditional approach with several learnable parameters. 
	In many cases, such structure is constructed by unfolding an iterative algorithm into a signal flow graph. 
	In other words, MDML focus on how to construct a learning structure or neural network with a similar process of traditional approaches to reduce the complexity of modeling. 
	Therefore, its has the advantage of reducing the complexity of modeling learning structures.
	On the contrary, the learning module in KDML is designed to fine-tune the outcome of traditional approaches. 
	Therefore, the network structure of learning module will be simplified which can be easily applied with no or less modifications to existing algorithms.

	\section{Case Study: Channel Estimation}
	\label{system}	
	In this section, a channel estimator considering both time and frequency selective fading is taken as the implementation of KDML. We introduce the system model at first and present the model implementation. 

	\subsection{Channel Model}
	We consider both time and frequency selective fading effects, which have been rarely investigated in existing works. The difficulty for this scenario is how to estimate the channel accurately without complex neural network structure. With Jakes fading model, in which $T$ plane waves are assumed to arrive in the uniform directions, the time selective fading can be given as\cite{1010873}
	\begin{equation}
		\label{jakes}
		g(t) = \frac{E_0}{\sqrt{2M+1}}\{g_I(t)+jg_Q(t)\},
	\end{equation}
    where $M=(T/2-1)/2$, $E_0$ is the average of the fading channels, and $g_I$ and $g_Q$ are defined as
	\begin{equation}
		\label{jakes2}
		\begin{aligned}
			&g_I(t) = 2\sum_{m=1}^{M}(\cos\phi_m\cos\omega_mt) + \sqrt{2}(\cos\phi_T\cos\omega_dt),\\
			&g_Q(t) = 2\sum_{m=1}^{M}(\sin\phi_m\cos\omega_mt) + \sqrt{2}(\sin\phi_T\cos\omega_dt),
		\end{aligned}
	\end{equation}
    where $\omega_d$ stands for the maximum Doppler shift and $\omega_m = \omega_d\cos(\frac{2\pi m}{T})$ for $m \in \{1,2,...,M\}$. We set the the initial phases as $\phi_T=0$ and $\phi_m = \frac{\pi m}{M+1}$.
	
	%The tapped delay line model \cite{279957} is shown in \eqref{tapped}
	The frequency selective fading can be expressed with the delay line model as \cite{279957}
	\begin{equation}
		\label{tapped}
		h(t) = \sum_{i=1}^{I}\sqrt{P_i}g_i(t)\delta(t-\tau_i)
	\end{equation}
	where $I$ is the number of distinguishable multi-paths, $P_i$ and $\tau_i$ stand for the power and time delay of the $i$th path and the multi-path component $g_i(t)$ is generated by the Jakes fading model.

	\subsection{Transmission Model}
	
	OFDM is an effective multiplexing technology to eliminate inter-symbol interference (ISI) and inter-carrier interference (ICI).
	Meanwhile, it can also effectively resist frequency selective fading.
	%The conceptual diagram of the OFDM system is shown in Fig. \ref{OFDM_frame}.
	At the transmitter, the transmitted data firstly passes a serial-to-parallel converter and is divided into blocks of size $N$ thereafter.
	Then the data blocks go through a modulator and are further allocated to subcarriers by an $N$-point IFFT. So the original data is seen as a frequency domain signal and becomes a time-domain signal after passing a parallel-to-serial converter.
	Before transmitting, a guard interval (GI) is added between adjacent OFDM symbols to minimize ISI and ICI. 
	The cyclic prefix is used as the guard interval, which is the same as tail data of an OFDM symbol.
	%In this paper, Cyclic Prefix (CP)\cite{372015} is used as the guard interval. CP is the same with tail data of the OFDM symbol.
	%Due to this cyclic structure, the subcarriers are orthogonal to each other. And when the length of CP is greater than $\max{\{\tau_i\}}$, it can eliminate the ISI.
	By digital-to-analog conversion, OFDM symbols will pass through an aforementioned multi-path fading channel. After these operations, the received data is given as
	\boldmath
	\begin{equation}
		y_{\rm{s}} = h_{\rm{s}} \cdot x_{\rm{s}} + w
	\end{equation}
	\unboldmath
	where $\bm{y_{\rm{s}}, x_{\rm{s}},h_{\rm{s}},w}$ stand for the receive data, transmitted data, FFT of $h(t)$ and AWGN, respectively.

	\subsection{Model Implementation}
	\subsubsection{Knowledge Module}
	Based on the knowledge of prior estimation, channel estimation can be performed by using pilots and there are two traditional algorithms for channel estimation, LS and MMSE.
	%Based on the KDML, we select one of these two algorithms to construct the knowledge module.
	%Compared with MMSE, LS has lower complexity at the cost of lower accuracy. 
	%KDML is supposed to be suitable for all kinds of domain knowledge, therefore both algorithms will be used to construct the knowledge module.
	By simply ignoring the Gaussian white noise, LS gives the channel estimation as
	
	\begin{equation}
	\label{LS}
		\hat{\bm{h}}_{\rm LS} = \frac{\bm{y_{\rm{s}}}}{\bm{x_{\rm{s}}}}.
	\end{equation}
	MMSE takes advantage of the autocorrelation of the channel to estimate the CSI\cite{CharlesMinimum} and aims at minimizing the mean square error between transmitted and received data as
	\begin{equation}
		\mathop{\arg\min}_{W} \ E\{(\bm{y_{\rm{s}}}-W\bm{x_{\rm{s}}})(\bm{y_{\rm{s}}}-W\bm{x_{\rm{s}}})^H\}.
	\end{equation}
	The solution can be given as
	\begin{equation}
		\hat{\bm{h}}_{\rm MMSE} = R_{\bm{h_{\rm{s}}h_{\rm{s}}}}(R_{\bm{h_{\rm{s}}h_{\rm{s}}}} + \frac{\sigma_{\bm{w}}^2}{\sigma_{\bm{x_{\rm{s}}}}^2}\bm{I})^{-1}\hat{\bm{h}}_{\rm LS}
	\end{equation}
%	R_{\bm{h_{\rm{s}}y_{\rm{s}}}}R_{\bm{y_{\rm{s}}y_{\rm{s}}}}^{-1}\bm{y_{\rm{s}}}
	where $\bm{I}$ denotes unity matrix, $(\cdot)^H$ means the Hermitian and $R_{\bm{h_sh_s}} = E\{\bm{h_sh_s}^H\}$ stands for autocorrelation of channel. Due to the lack of real CSI, it is approximately calculated as $R_{\bm{h_sh_s}} \approx E\{\hat{\bm{h}}_{\rm LS}\hat{\bm{h}}_{\rm LS}^H\}$	
	For pilots, CSI can be calculated by these two algorithms directly. And for other data, CSI is obtained by linear interpolation. Suppose two adjacent pilot positions are $l_1 $ and $l_2$ ($l_1 \le l_2$), for any data located in $i \in [l_1, l_2]$, the interpolation is
	\begin{equation}
		\hat{\bm{h}}_{i} = \hat{\bm{h}}_{l_1,est} + \frac{\hat{\bm{h}}_{l_2,est} - \hat{\bm{h}}_{l_1,est}}{l_2 - l_1} \cdot (i-l_1).
	\end{equation}
	\subsubsection{Learning Module}
	\label{LSTM_Str}
	
	\begin{figure}[!t]	
		\centering
		\includegraphics[width=0.4\textwidth]{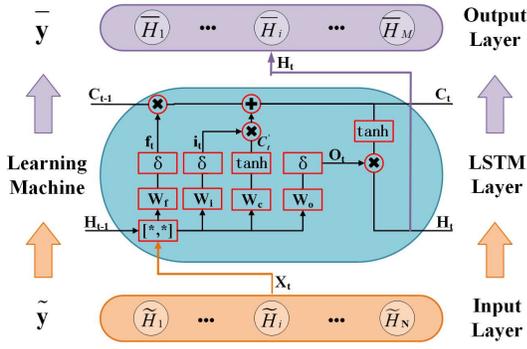}
		\caption{ Structure of Learning Module %\comm{any difference with existing one? source?}
		}
		\label{LSTM_cell}
	\end{figure}
    Let $\bm{\widetilde{H}}$ and $\bm{\overline{H}}$ denote the channel estimation of traditional algorithms and outcome of learning module, respectively.
	After obtaining $\bm{\widetilde{H}}$, the learning module is designed to fine-tune $\bm{\widetilde{H}}$ by mining the hidden information in the time domain. 
    %$\bm{\widetilde{H}}$ denotes the output of the learning module and consists of CSIs of both the positions of pilots and data.
	%The learning module will treat $\bm{\widetilde{H}}$ as a time series, which can be well analyzed by the recurrent neural network (RNN).
	The whole structure of the learning module is shown in Fig.\ref{LSTM_cell}.
	At the input layer, $\bm{\widetilde{H}}$ is firstly transformed into a dataset of supervised learning. %The element representing time $\bm{t}$ in the dataset contains features and labels.
	For this dataset, features are composed of rough estimations at different moments, and $\bm{N}$ stands for time-steps, which indicates how many previous CSI are used to fine-tune the channel estimation. Labels include $\bm{M}$ estimations at different moments.
	The instance in the dataset is given as:
	$\bm{x=[\widetilde{H}_{t-N}, ..., \widetilde{H}_{t-1}, \widetilde{H}_{t}] \in \mathbb{C}^{1*N};
		y =[\widetilde{H}_{t+1}, \widetilde{H}_{t+2}, ... , \widetilde{H}_{t+M}]} \in \mathbb{C}^{1*M}$
	, where $\bm{\widetilde{H}_{t}}$ denotes the channel estimation based on traditional algorithms at time
	$\bm{t}$. Because the neural network only accepts real numbers, $\bm{\widetilde{H}_{t}}$ will be split into real and imaginary two parts.
	
	The second layer consists of LSTM, which is a variation of RNN.
	Although there is only one LSTM cell presented in Fig.\ref{LSTM_cell}, it will be recycled many times during the training process.
	%\revise{LSTM\cite{HochreiterLong} is a variation of RNN}
	%LSTM cell is the basic cycle unit of an LSTM layer.
	To avoid gradient vanishing or exploding, there are three essential components in the LSTM cell, i.e, Forget Gate, Input Gate, and Output Gate.
	As Fig. \ref{LSTM_cell} presents, the LSTM cell will perform following operations
	\begin{subequations}
		\label{LSTM_equ}
		\begin{align}
			&\text{Forget Gate: \ } f_t = \delta([X_t,H_{t-1}]\cdot W_f + b_f ) \\
			&\text{Input Gate: \ }
			\begin{cases}
				i_t&= \delta([X_t,H_{t-1}]\cdot W_i +  b_i)  \\
				C_t^{'}&= \tanh ([X_t,H_{t-1}]\cdot W_c +  b_c)\\
				C_t&= f_t \times C_{t-1} + i_t \times C_t^{'}
			\end{cases}                      \\
			&\text{Output Gate: \ }
			\begin{cases}
				O_t&= \delta([X_t,H_{t-1}] \cdot W_o+  b_o)  \\
				H_t&= O_t \times \tanh(C_t)
			\end{cases}
		\end{align}
	\end{subequations}
	where $\delta$(·) is the Sigmod function,
	$\mathit{S(x)}= \frac{1}{1 + e^{-x}}$ and $tanh$(·) is hyperbolic tangent function.
	$X_t,H_t$ and $C_t$ stand for the input data, hidden state and cell state of LSTM at time $t$, respectively.
	Forget Gate combines the input data and previous hidden state, which decides what information to be retained and what to be discarded.
	Input Gate is used to update the cell state. It calculates what percentage of information to be updated at first.
	Then, the cell state will be updated by mixing the previous and candidate cell state.
	Output Gate designated to decide what is to be output and $H_t$ is both the new hidden state and output of the LSTM cell. %During the training process of LSTM, weights ($W_*$) and biases ($b_*$) will be learned.
	
	The output layer is made of the fully connected layer whose function can be seen as:
	\begin{equation}
		g(\bm{x}) = W_g\bm{x}+b
	\end{equation}	
	After the above operations, the final output of LSTM network is $\bm{\overline{H}} = \{\bm{\overline{H}_{t+1}}, ..., \bm{\overline{H}_{t+M}} \}$.
	Then the loss function is defined as:
	\begin{equation}
		Loss = \frac{1}{M} \sum_{i=1}^{M} \left\|\bm{\overline{H}_{t+i}} - \bm{\widetilde{H}_{t+i}}\right\|^2
		\label{output}
	\end{equation}
	To improve the feasibility, even during the training process, only $\bm{\widetilde{H}}$ is utilized rather than the real CSI.
	All parameters will be updated by the optimization algorithm, Adam\cite{Kingma2014Adam}.
	%And the optimization algorithm is Adam algorithm\cite{Kingma2014Adam}.
%	After offline training, $\bm{\mathbf{\Theta}}$ is saved and then loaded to test the performance over test set .

	\subsection{Time Complexity Analysis}
	\label{complexity}
	
	There are two kinds of complexity about neural network, space and time complexity. 
	%The former describes the capacity of a certain model. The number of parameters is often considered as spatial complexity. 
	The former describes the capacity of a certain model which is considered as the number of parameters.
	The latter is closely connected to hardware run time, which is often represented by the number of Floating Point Operations (FLOPs). Considering that the communication system is strict with time-delay, it is necessary to analyze the time complexity of the proposed estimator.
	
	According to Eq.\eqref{LSTM_equ}, it is obvious that the number of FLOPs is mainly affected by four group parameters. More specifically, the number of parameters of the four groups is the same. Denote the size of input data, hidden size and the size of output as $i$, $m$, and $l$.
	Then we have $H_t \in \mathbb{R}^{1*m}, W_f \in \mathbb{R}^{(i+m)*m}, W_g \in \mathbb{R}^{m*l}$. And all weight matrices in LSTM layer have the same size. Therefore, the required FLOPs of LSTM layer is $4 * (i*m + m*m + m)$ , and $m * l$ for output layer. Assume the length of transmitted data is $n$, then the total required FLOPs is $n * 4 * (i*m + m*m + m) + n*m*l$.
	Without loss of generality, $m \gg i $ and $m \gg l$. Thus, the time complexity of LSTM is given as,
	\begin{equation}
		T_{LSTM}(n) = O(n*m^2)
	\end{equation}
	Meanwhile, it is easy to figure out that the time complexity of the LS algorithm is $O(n)$. And for the MMSE algorithm, the most complex operations are matrix inversion, whose time complexity is $O(n^3)$. The complexity of the LS, LSTM is proportional to $n$, which is far less than MMSE.
	
%	Given the above analysis, if LS is chosen to get rough channel estimation, for the proposed channel estimator, its time complexity is still proportional to $n$.  Although the time complexity is also proportional to the square of $m$, with the increasing of data length, its influence will diminish.
	
	\section{Experiments Design and Results}
%\cite{8043424} 
	The OFDM system parameters and the major parameters for LSTM are listed in TABLE \ref{sys_para} and \ref{lstm_para}, respectively.
	For brevity, we denote the reciprocal of the pilot density as number of pilot spacing(NPS) in the following experiments.
	In the simulation, the bit sequence is randomly generated and the number of multi-paths is three with each having a random and independent different delay and maximum Doppler shift.
	The channel estimation dataset has a training set of 27,000 examples and a testing set of 3,000 examples, a total of 30,000 examples in 5 types which contains data with a certain SNR.
	
	Moreover, it is worth noting that both the theoretical and simulation results of the MMSE estimation are considered, which are represented by MMSE and MMSE-Sim, respectively.
	And we use estimations of LS, MMSE, and real CSI to construct three channel estimators, which are named as KDML(LS), KDML(MMSE), and KDML(H), respectively.
	Besides, we also train a Multilayer Perceptron(MLP) network as the performance baseline of ML, which contains six full connected layers with ReLu activation function and has the same dataset as KDML.

	\begin{table}[t]
		\centering
		\caption{Communication system parameters}
		\setlength{\tabcolsep}{4mm}
		\begin{tabular}{|l|c|c|c|c|}\hline
			Sub-carrier interval&\multicolumn{4}{|c|}{15KHz}\\ 		\hline
			Sampling rate& \multicolumn{4}{|c|}{15.36MHz} \\ 		\hline
			IFFT Size& \multicolumn{4}{|c|}{1024}\\   \hline
			Modulate&   \multicolumn{4}{|c|}{QPSK}\\             \hline
			Traditional Algorithm&\multicolumn{2}{|c|}{LS}&\multicolumn{2}{|c|}{MMSE} \\ 		\hline
			NPS&2&4&8&16\\ 		\hline
		\end{tabular}
		\label{sys_para}
	\end{table}
	
	\begin{table}[t]
		\centering
		\caption{Learning module parameters}
		\setlength{\tabcolsep}{4mm}
		\begin{tabular}{|l|c|}\hline
			Net architecture&{LSTM+Full Connected Layer}\\ 		\hline
			Activation function& Relu \& Tanh \\ 		\hline
			Loss function& {MSE} \\ 		\hline
			Optimizer&{Adam}\\ 		\hline
			Hidden size&{128}\\ \hline
			Learning rate&{0.01} \\ \hline
			Batch size&{500} \\ \hline
			Training number&{27000} \\ \hline
			Test number&{3000} \\ \hline	
			Epoches&{100} \\ \hline
		\end{tabular}
		\label{lstm_para}
	\end{table}
	%\comm{It is worth mentioning that in the experiments two different MMSE results are considered, MMSE-Sim and MMSE. The difference between these two methods is $R_{HH}$. For MMSE-Sim, $R_{HH} = E\{\hat{H}_{LS}\hat{H}_{LS}^H\}$ while $R_{HH} = E\{{H}{H}^H\}$ for MMSE. }
	
	\subsection{Performance Comparison with Other Algorithms}
	
\begin{figure}[!t]	
		\centering
		\includegraphics[width=0.45\textwidth]{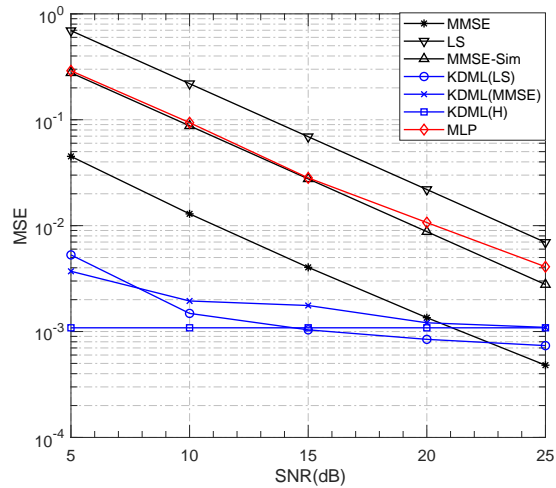}
		\caption{ The mean square error of channel estimation(LS, MMSE, MLP, KDML) (Pilot density = 50\%).}
		\label{MSE_NPS=2}
	\end{figure}

	Fig.\ref{MSE_NPS=2} presents the MSE performance of all five methods when NPS = 2.

	Firstly, LS has the worst performance due to ignoring noise and there is a significant gap between MMSE-Sim and LS, which meets our common understandings.
	Meanwhile, the theoretical performance of MMSE is one order of magnitude better than that of LS.
	%Then, to evaluate the performance of the proposed estimator, it is supposed to be divided into two intervals by the SNR. And the division is based on whether the SNR is greater than 20dB. 
	For the low SNR($\leq$20dB) interval, the MSE improvement of the proposed estimator can achieve over two orders of magnitude than that of LS, which is even better than that of MMSE.
	But the gap between KDML estimator and LS diminishes gradually. A possible reason is that as the SNR increases, there are less and less noise hidden in the estimation of LS.
	When it comes to the high SNR interval, MMSE has better performance than KDML instead. KDML(H), which takes the real CSI as knowledge is simulated to explain this phenomenon. 
	Because the real CSI is not affected by SNR, its MSE performance basically does not change with the SNR, which can be regarded as inherent error of the learning module we used. 
	We speculate that this error may gradually become the dominant factor affecting the performance of KDML with increasing SNR. So that, KDML estimator will lose its performance advantage when SNR is greater than 20dB.
	Even so, how to improve the performance during low SNR interval is an important issue for ML based estimator.
	The proposed KDML estimator gives a preliminary solution of this issue. 
	In general, such results prove the feasibility of KDML when SNR in 5$\sim$25dB.
	 
	 Secondly, the performance of MLP is better than that of LS, but there is still a significant gap between MLP and KDML. 
	 The reason we choose MLP as traditional machine learning model as benchmark is that, MLP is one of the most widely used structures, which has already been used for channel estimation.
	 Although MLP can not stands for the most advanced ML algorithms, there are still reasons to believe that the addition of knowledge module brings certain performance gain compared to pure machine learning solutions.

%	\subsection{Performance Comparison with Different Knowledge}
	Finally, whether better knowledge module output can bring better KDML performance is a crucial issue.
	To address this issue, experiments with different rough channel estimations are also shown in Fig.\ref{MSE_NPS=2}.
	Here, the performance of KDML(LS), KDML(MMSE) and KDML(H) reveals one possible interesting insight: better input of learning module doesn't necessarily lead to better performance in KDML.
	KDML(H) is designed as the baseline of KDML with different knowledge, however, counter-intuitively, as the SNR increases, KDML(LS) performs better than KDML(H) instead. 
	Such phenomenon implies that there may also exist inherent error between the learning model output and the real LS channel estimation when training KDML(LS). 
	Eq.\eqref{output} means that the learning module in KDML is supposed to recover CSI from the noisy channel estimations, while the inherent error may be able to offset the noise information, thus KDML(LS) can perform better than KDML(H). 
%	A similar principle can also be found in a well-known method of preventing over-fitting, early stop.
	Similarly, although the channel estimation of MMSE is better than that of LS, KDML(MMSE) perform worse than KDML(LS) when the SNR is greater than 10dB. 
	Here how to apply the channel auto-correlation is quite different. 
	On the one hand, for the KDML(LS), the channel auto-correlation is only mined by the LSTM network as a kind of time-series correlation. 
	On the other hand, for the KDML(MMSE), the channel auto-correlation is firstly calculated in the MMSE algorithm, thereafter mined by the LSTM network. 
	Thus, the difficulty of extracting correlations by the LSTM network increase when using the estimation of MMSE as the input of LSTM. 
	And it may eventually causes the performance of KDML(LS) is better than KDML(MMSE).

	\subsection{Performance Comparison with Different Pilot Density}	
	\begin{figure}[!t]	
		\centering
		\includegraphics[width=0.44\textwidth]{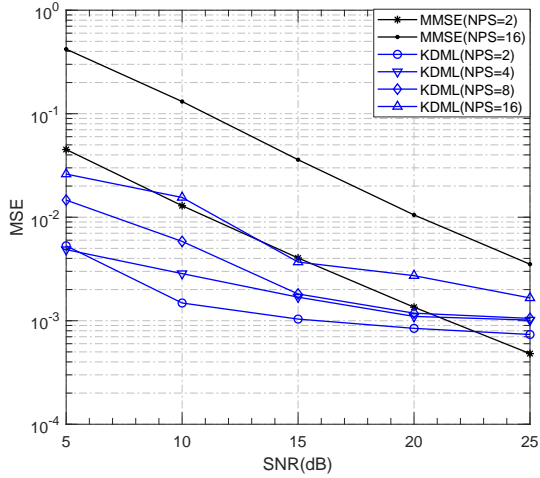}
		\caption{ The mean square error with different pilot density.}
		\label{LSTM_MMSE_Compare}
	\end{figure}
	Fig.\ref{LSTM_MMSE_Compare} demonstrates the performance of KDML with different pilot density. 
	At first, as NPS increases from 2 to 16, the performance of the proposed estimator slightly decreases. 
	Compared with MMSE, the proposed KDML estimator is less affected by the pilot density. 
	This phenomenon demonstrates the robustness of the proposed estimator, which can be applied with different pilot density with little performance loss.
	Furthermore, when the SNR is lower than 20dB, the performance of KDML(LS) is not only higher than that of MMSE(NPS=16) but also very close to that of MMSE(NPS=2). 
	In other words, under the same performance, the proposed estimator can save nearly 75\% pilot resources comparing with the MMSE estimator. 
	This characteristic indicates that the proposed estimator has great potential in the massive MIMO scene.
	
	\section{Conclusion}
	In this paper, we proposed the knowledge-driven machine learning model which combines the power of knowledge and machine learning.. 
	KDML can significantly reduce training costs and simplify network structure. 
	Experiment results demonstrate the efficiency of the proposed estimator, which outperforms the traditional algorithms or conventional ML and exhibits robustness with pilot density and potential for saving pilot resources.
%	This implementation proves the feasibility and reliability of the proposed KDML.

	\section*{acknowledgment}
	This work was partially supported by Key Program of Natural Science Foundation of China under Grant(61631018), Huawei Technology Innovative Research(YBN2018095087).

\bibliographystyle{IEEEtran}
\bibliography{ref}

\end{document}